\documentclass[graybox]{svmult}

\usepackage{mathptmx}       
\usepackage{helvet}         
\usepackage{courier}        
\usepackage{type1cm}        
\usepackage{makeidx}         
\usepackage{graphicx}        
\usepackage{multicol}        
\usepackage[bottom]{footmisc}

\usepackage{amsmath}
\usepackage{amssymb}
\usepackage{amsfonts}
\usepackage{algorithm, algorithmic}

\makeindex             

\begin{document}

\title*{PAWL-Forced Simulated Tempering}
\author{Luke Bornn}
\institute{Luke Bornn \at Harvard University, 1 Oxford St., Cambridge, MA, 02138 USA\\\email{bornn@stat.harvard.edu}}
%
%
\maketitle

\vspace{-85px}

\abstract{In this short note, we show how the parallel adaptive Wang-Landau (PAWL) algorithm of Bornn et al. (2013) can be used to automate and improve simulated tempering algorithms.  While Wang-Landau and other stochastic approximation methods have frequently been applied within the simulated tempering framework, this note demonstrates through a simple example the additional improvements brought about by parallelization, adaptive proposals and automated bin splitting.}

\section{A Parallel Adaptive Wang-Landau Algorithm}
\label{sec:1}
The central idea underlying Wang-Landau (\cite{wang1992}) and related algorithms is that instead of generating samples from a target density $\pi$, it is sometimes more efficient to instead sample a strategically biased density $\tilde{\pi}$.  In the case of Wang-Landau, the goal is to sample
\begin{align}
  \tilde{\pi}(x) = \pi(x) \times \frac{1}{d} \sum_{i=1}^{d} \frac{\mathcal{I}_{\mathcal{X}_i}(x)}{\int_{\mathcal{X}_i} \pi(x) \mathrm{d}x}
	\label{BiasedDensity}
\end{align}
where $\mathcal{I}_{\mathcal{X}_i}(x)$ is equal to $1$ if $x\in\mathcal{X}_i$ and 0
otherwise.  Interestingly, this biased target ensures each of the partitions of the space $(\mathcal{X}_i)_{i=1}^d$ are visited equally: $\int_{\mathcal{X}_i}\tilde{\pi}(x) \mathrm{d}x = \int_{\mathcal{X}_j}\tilde{\pi}(x)\mathrm{d}x, \; \forall i, j \in (1,\dots,d)$. Additionally, the restriction of the modified distribution $\tilde{\pi}$ to each set
$\mathcal{X}_i$ coincides with the restriction of the target distribution $\pi$
to this set up to a multiplicative constant; namely for all $i$, $\tilde{\pi}(x) \propto \pi(x), \; \forall x\in \mathcal{X}_i$.

While the biased density $\tilde{\pi}(x)$ has desirable properties, an obvious problem is that calculating $\int_{\mathcal{X}_i} \pi(x) \mathrm{d}x$ is not straightforward.  As such, the Wang-Landau algorithm creates estimates $\theta_t$ of these quantities at each step $t$.  Algorithm 1 provides psuedo-code for the algorithm.
\begin{algorithm}
\caption{Simplified Wang-Landau Algorithm\label{algo:WL}}
\begin{algorithmic}[1]
\STATE Partition the state space into $d$ regions $\{ \mathcal{X}_1, \dots,
\mathcal{X}_d \}$ along a reaction coordinate $\xi(x)$. 
\STATE First, $\forall i\in \{ 1,\dots, d \}$ set $\theta(i) \leftarrow 1$.
\STATE Choose a decreasing sequence $\{ \gamma_t \}$, typically $\gamma_t = 1/t$.
\STATE Sample $X_0$ from an initial distribution $\pi_0$.
\FOR {$t=1$ to $T$}
      \STATE Sample $X_t$ from $P_{\theta_{t-1}}(X_{t-1},\cdot)$, a
transition kernel with invariant distribution $\tilde{\pi}_{\theta_{t-1}}(x)$.
      \STATE Update the bias: $\log \theta_{t}(i) \leftarrow \log  \theta_{t-1}(i) + \gamma_t
(\mathcal{I}_{\mathcal{X}_i}(X_t) - d^{-1})$.
      \STATE Normalize the bias: $\theta_t(i) \leftarrow \theta_t(i) / \sum_{i = 1}^d \theta_t(i)$.
\ENDFOR
\end{algorithmic}
\end{algorithm}
In the full version of the algorithm, the step size $\gamma_t$ is only reduced when all of the regions $\{ \mathcal{X}_1, \dots,\mathcal{X}_d \}$ have been uniformly explored as measured by the flat histogram criterion $\max_{i \in [1,d]} \vert \nu(i) - d^{-1}\vert < c/d$ where $\nu(i)$ is the proportion of samples within $\mathcal{X}_i$ since the last time the flat histogram criterion was met.  Here $c$ is a user-specified threshold.  The reader is referred to \cite{bornn2013} for a full description and discussion of the algorithm, as well as details on stabilizing the algorithm through parallelization, introducing adaptive proposals, and automating the partitioning of the space.  These three improvements, applied to simulated tempering, will be the focus of this work.

\section{Simulated Tempering}
The use of stochastic approximation algorithms, including Wang-Landau, within simulated tempering has been suggested by various authors (see, for example, \cite{geyer1995} and \cite{atchade2010}).  In this note, we further examine the improvements proposed in \cite{bornn2013}, namely parallelization, adaptive proposals, and automatic partitioning of the space.  The primary idea of simulated tempering is to sample from a tempered distribution $\pi_T(x) = \pi(x)^{1/T}$ for some temperature $T$.  The algorithm proceeds by setting a temperature ladder $T=1,\dots, T_{max}$ and running a Markov chain on the pair $(x,T)$.  As such, the chain explores the state space $\mathcal{X}$ while moving up and down the temperature ladder.  Readers are referred to \cite{marinari1992, geyer1995} for further details.  Of note for our purposes, however, is that one is able to specify pseudo-priors on the different steps of the ladder to ensure equal occupation numbers -- time spent in each step of the ladder -- which is a task well-suited for stochastic approximation.

\begin{figure}
\centering
\includegraphics[width=0.95\textwidth]{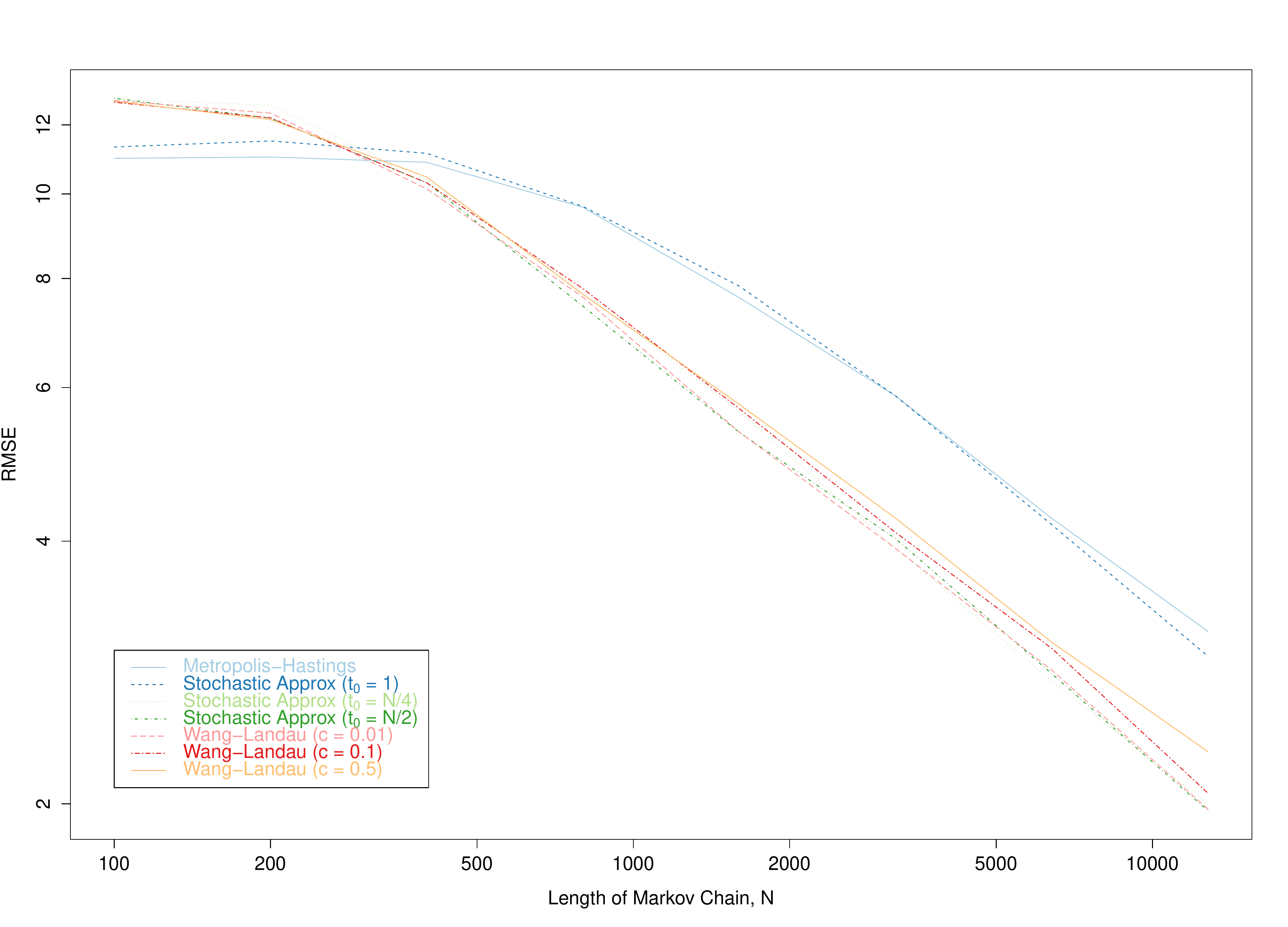}
\caption{\label{fig:step_sizes} RMSE for estimating the mean in the bimodal density for various simulated tempering configurations.  We see that Wang-Landau (provided $c$ is small) and stochastic approximation with deterministic step size decreases (provided $t_0$ is large) both perform well.}
\end{figure}

To test these (potential) improvements to simulated tempering, we employ a small bimodal density.  Specific, we set $\pi(x)$ to be an equally-weighted mixture of two standard normal distributions, one centered at $-15$ and the other at $15$.  As such, the distribution has two modes (at $x=-15$ and $x=15$) with a large low-density valley separating them. As a result, estimating the mean ($0$) is a natural challenge for any sampler.  We run $1000$ chains each of length $N$ (for various $N$), and calculate the root mean squared error (RMSE) between the posterior mean (calculated from all states with $T=1$) and the true mean of $0$.  We compare standard simulated tempering using Metropolis-Hastings with uniform pseudo-priors (using a Gaussian random walk with standard deviation 10, and temperatures $T=1,2,\dots,9,10$) to that using stochastic approximation adjusted such that the pseudo-priors ensure equal occupation numbers.  See \cite{atchade2010} for details.  We use standard stochastic approximation with step sizes $\gamma_t = t_0 / \text{max}(t_0,t)$ for $t_0=1, N/4, N/2$.  In other words, the step size starts decreasing after 1 iteration, $N/4$ iterations, or $N/2$ iterations, respectively.  We also explore Wang-Landau, which automatically decreases the step size after a flat histogram criterion is met. We look at $3$ values of the user-specified tuning parameter $c$, namely $c=0.01,0.1,0.5$.  Figure \ref{fig:step_sizes} displays the RMSE as a function of $N$ for each algorithm.  We see that all of the stochastic approximation algorithms (including Wang-Landau) perform similarly in this simple example. It has been argued, however, that in more complex situations Wang-Landau will outperform stochastic approximation with deterministicly decreasing step size \cite{atchade2010}.

\begin{figure}
\centering
\includegraphics[width=0.95\textwidth]{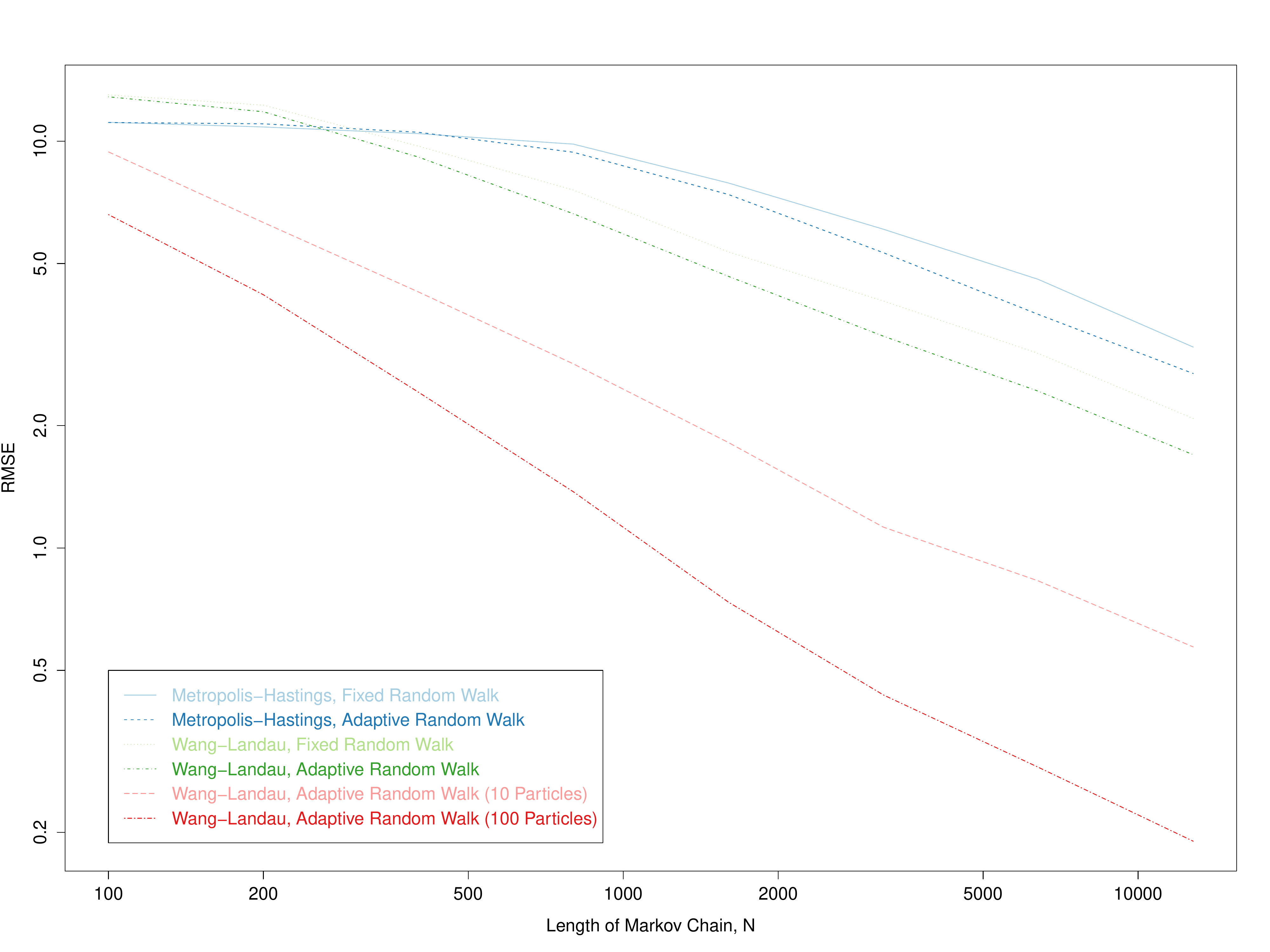}
\caption{\label{fig:adaptive} RMSE for estimating the mean in the bimodal density for various simulated tempering configurations with and without adaptive proposals.}
\end{figure}

In Figure \ref{fig:adaptive} we similarly compare the simple Metropolis-Hastings simulated tempering algorithm to the Wang-Landau version (using $c=0.1$) with and without adapting the proposal standard deviation (set to target an acceptance ratio of $0.234$); see \cite{bornn2013} for specifics.
It is clear that adaptation in the proposal mechanism provides significant gains to both the standard simulated tempering algorithm as well as the Wang-Landau version.  Further improvements might be made by considering mixture proposals tailored to each step on the temperature ladder, rather than being optimized to create a given acceptance rate across all temperatures.  Figure \ref{fig:adaptive} also displays the adaptive Wang-Landau algorithm in parallel with $10$ and $100$ particles, demonstrating vastly improved convergence of the algorithm. With $M$ particles, the approximate improvement in RMSE is $\sqrt{M}$, which is roughly equivalent to if we were to run a single chain for $M \times N$ iterations.  However, due to vectorization the parallel version does not take $M$ times as long to run.  In our examples, $M=10$ and $M=100$ particles took $1.8$ and $6.2$ times longer than the single chain, respectively.

We also explored automatic setting of the temperature ladder using the bin-splitting method proposed in \cite{bornn2013} (not shown).  However, in this small example the advanced binning method performed similarly to simply fixing the temperature ladder to the integers $1,\dots,10$. We suspect that in more complicated settings where the results are more sensitive to the temperature ladder the automatic binning approach will bring additional benefit.

\section{Conclusion}
This brief note has employed a simple bimodal example to demonstrate the benefits of embedding adaptive proposals, parallelization, and automatic bin splitting within the simulated tempering framework. Due to space limitations, many pertinent references and ideas have been excluded, though the interested reader might follow the citation trail to further explore these algorithms.  If there is a single takeaway, it is that sometimes ``stacking'' multiple computational techniques can lead to significant improvements in performance.  In this case, parallelization and adaptive proposals provide signifant improvements to simulated tempering with the Wang-Landau algorithm; additionally, they are straightforward to implement through the R package \texttt{PAWL}, available online.

Ongoing work involves applying these simulated tempering methods to learn latent dimensions in nonstationary spatial models (\cite{bornn2012}), which due to partial identifiability of the parameter space show particular promise for benefiting from the ideas presented herein.  Specifically, as this class of models is new and as-yet poorly understood, it is unclear apriori how to determine the scale of the proposal distribution as well as set the temperature ladder.

%
%
%

\end{document}